\documentclass[twocolumn,showpacs,preprintnumbers,amsmath,amssymb,prl]{revtex4}

\usepackage{graphicx}
\usepackage{dcolumn}
\usepackage{bm}

\begin{document}

\input epsf.sty

\draft
\widetext

\title
{
Magnetic Field Effect on Static Antiferromagnetism in Electron-Doped Superconductor Pr$_{1-x}$LaCe$_{x}$CuO$_{4}$ ($x$=0.11 and 0.15)
}


\author{M. Fujita$^{1}$}%
\email{fujita@imr.tohoku.ac.jp}

\author{M. Matsuda$^{2}$}

\author{S. Katano$^{2}$}

\author{K. Yamada$^{1}$}%

\affiliation{%
$^{1}$Institute for Material Research, Tohoku University, Katahira, Sendai 980-8577, Japan
}%

\affiliation{%
$^{2}$Advanced Science Research Center, Japan Atomic Energy Research Institute, Tokai, Ibaraki 319-1195, Japan
}

\date{\today}



\begin{abstract}

Effects of magnetic fields (applied along the c-axis) on static spin correlation were studied for the electron-doped superconductors Pr$_{1-x}$LaCe$_{x}$CuO$_4$ with $x$=0.11 ($T_c$=26 K) and $x$=0.15 ($T_c$=16 K) by neutron scattering measurements. 
In the $x$=0.11 sample located near the antiferromagnetic(AF) and superconducting phase boundary, the field dependence of both the magnetic intensity at T=3 K and the onset temperature of the magnetic order exhibits a peak at $\sim$5 T. 
In contrast, in the overdoped $x$=0.15 sample a static AF order is neither observed at zero-field nor induced by the field up to 8.5 T. 
Difference and similarity of the field effect between the hole- and electron-doped high-$T_c$ cuprates are discussed. 

\end{abstract}


\pacs{74.72.Jt, 74.25.Ha, 75.50.Ee, 61.12.Ld} 

\maketitle

Magnetism in lamellar copper oxides is widely believed to play an important role in the mechanism of high-$T_c$ superconductivity\cite{Kastner}. 
Extensive neutron scattering measurements have indeed shown an intimate relation between incommensurate (IC) low-energy spin fluctuations observed in the hole-doped ($p$-type) La$_{2-x}$Sr$_{x}$CuO$_{4}$ (LSCO)\cite{Yoshizawa,Birgeneau,Cheong} and their superconductivity\cite{Yamada_LSCO}. 
Recentlly, neutron scattering study on the superconducting (SC) LSCO with $x$$\sim$1/8 and excess-oxygen doped La$_2$CuO$_{4+y}$ revealed an enhancement of the long-ranged IC magnetic order by magnetic fields applied along the c-axis \cite{Katano,Lake_Nature,Khaykovich1,Khaykovich2}. 
Lake {\it et al}., furthermore, found field-induced slow spin fluctuations below a spin-gap energy, which shows a tendency toward the magnetic order, in the optimally doped LSCO\cite{Lake_Science}. 
These authors discussed that the antiferromagnetic (AF) insulator with IC correlations is a possible ground state after vanishing the superconductivity, as supported by theoretical studies\cite{Zhang,Ogata,Ichioka}. 
Thus, in order to clarify the universal nature of magnetism hidden behind the superconductivity, it is necessary to investigate whether the AF order is commonly observed by suppressing the superconductivity.

Important challenges have been made on the prototypical electron-doped ($n$-type) system of Nd$_{2-x}$Ce$_x$CuO$_4$ (NCCO) which shows commensurate spin fluctuations at the tetragonal (1/2 1/2 0) reciprocal-lattice position in both AF and SC phases\cite{Yamada_NCCO}. 
In Nd$_{1.86}$Ce$_{0.14}$CuO$_{4}$, at least down to 15 K Matsuda {\it et al}. observed no field effect on the AF order which coexists or phase separates with the bulk superconductivity\cite{Matsuda}
, whereas Kang {\it et al}. subsequently reported a field-enhanced huge magnetic intensity for Nd$_{1.85}$Ce$_{0.15}$CuO$_4$ and asserted the AF order as a competing ground state with the superconductivity, irrespective of carrier type\cite{Kang}. 
This discrepancy could be originated from  drastic(first-order like) doping dependence of the static magnetism near the boundary between AF and SC phases\cite{Uefuji,Uefuji2}. 
However, since the bulk superconductivity with identical $T_c$ of 26 K as well as the AF order appears in both samples, field-effects on the AF order and the superconductivity are still controversial. 
Furthermore, a effect on Nd$^{3+}$ spins is expected to be significant in NCCO, and therefore, the results may conceal the inherent field dependence of Cu$^{2+}$ spin correlations. 
Thus, a more comprehensive study on the $n$-type system is required using a sample with small rare-earth magnetic moments.

In this paper, we report the result of neutron scattering measurements performed under magnetic fields on the $n$-type Pr$_{1-x}$LaCe$_{x}$CuO$_4$ (PLCCO) with $x$=0.11 ($T_c$=26 K) and 0.15 ($T_c$=16 K), in which the effect of rare-earth moments is negligibly small compared with that in NCCO\cite{Fujita_mSR,Kadono}. 
Samples with $x$=0.11 and 0.15 are located in the vicinity of the phase boundary between AF and SC phases and in the overdoped SC phase, respectively\cite{Garcia,Fujita_mSR}. 
At zero field, the former sample shows a short-ranged commensurate AF order, while no evidence of a magnetic order was observed in the latter sample. 
The intrinsic field-effect on the AF order is, hence, expected to be clarified by the comparative study of these samples. 
In the $x$=0.11 sample, the magnetic field up to 5 T along the c-axis enhances both the magnetic intensity of elastic scattering at ($\pi$,$\pi$) and the ordering temperature ($T_m$). 
Further applying the field both intensity and $T_m$ start to decrease and under the field of 9 T these values nearly return back to those in the zero field. 
In contrast, no static order was induced in the $x$=0.15 sample in the field up to 8.5 T. 
Thus, analogous to the results for LSCO, magnetic field affects the static spin correlation existing at zero-field, but does not induce an AF order at least up fo 8.5 T. 
We newly observed a similar field dependence between the enhanced magnetic intensity and $T_m$.

Single crystals of PLCCO with $x$=0.11 and 0.15 were grown using a traveling-solvent floating-zone method. 
We annealed the as-grown single crystalline rods with $\sim$30 mm in length and 6 mm in diameter under argon gas flow at 960$^{\circ}$C for 12 hours followed by further annealing under oxygen gas-flow at 500 $^{\circ}$C for 10 hours. 
For the present samples, the amount of the oxygen atoms removed by the heat treatment was determined to be 0.05 per unit formula from the weight loss of the sample after treatment. 
\linebreak
\begin{figure}[t]
\centerline{\epsfxsize=3.1in\epsfbox{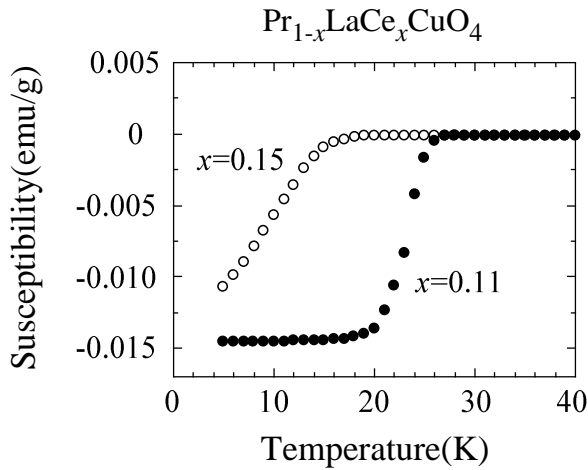}}
\caption
{
Magnetic susceptibility measured for single the crystals of Pr$_{0.89}$LaCe$_{0.11}$CuO$_4$ and Pr$_{0.85}$LaCe$_{0.15}$CuO$_4$ after the zero-field-cooling process. 
}
\end{figure}
%
\noindent
The edge of the annealed crystals was cut into a plate to characterize the superconductivity by magnetic susceptibility measurements. 
As shown in Fig. 1, both $x$=0.11 and 0.15 samples exhibit bulk superconductivity with $T_c$(onset) of 26 K and 16 K, respectively. 

Using the main part of crystals, elastic neutron scattering measurements were performed on the triple-axis spectrometer TAS-2 installed at the thermal neutron guide of JRR-3M of Japan Atomic Energy Research Institute (JAERI). 
We selected the incident neutron energy $E_i$ of 14.7 meV with the collimation sequence of 17$^{\prime}$-20$^{\prime}$-sample-20$^{\prime}$-80$^{\prime}$ (or 17$^{\prime}$-80$^{\prime}$-sample-40$^{\prime}$-80$^{\prime}$) for the investigation of the peak-profile with a high-resolution (or the peak-intensity against the temperature and the magnetic field). 
A pyrolytic graphite filter was placed before the sample to eliminate the higher-order harmonics of the incident neutrons. 
We mounted each single crystal with the CuO$_2$ sheets in the horizontal scattering plane and applied magnetic fields vertically to this scattering plane using a split-type cryocooled superconducting magnet. 
Crystallographic indexes are denoted as ($h$ $k$ 0) in the tetragonal notation with the reciprocal-lattice unit (r.l.u.) at 3 K of 1.579 $\AA$$^{-1}$, corresponding to the lattice constant of 3.979 $\AA$ along he Cu-O bonding. 
All measurements were done in the field cooling process after the field was set at above $T_c$($B$=0) and $T_m$($B$).

The spatial static spin correlations were studied by scans along [1 -1 0] direction through the AF zone center of (1.5 0.5 0) in the as-grown samples\cite{Fujita_neutron}. 
Figure 2 shows the scan-profiles at 3 K after subtracting the background at high temperatures. 
Intensities for both samples are normalized by their volumes. 
At low temperatures in zero-field, a weak intensity due to the magnetic order appears in the $x$=0.11 sample at a commensurate (1.5 0.5 0) position. 
\linebreak
%
%
\begin{figure}[t]
\centerline{\epsfxsize=2.45in\epsfbox{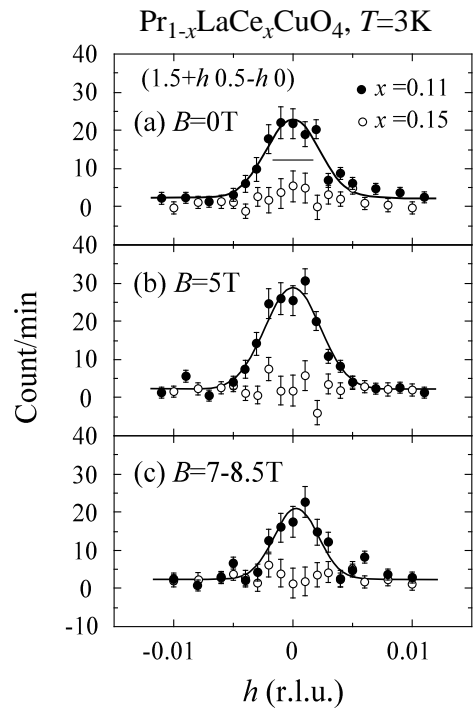}}
\caption
{
Background-subtracted peak profiles for Pr$_{1-x}$LaCe$_{x}$CuO$_4$ with $x$=0.11(closed circles) and $x$=0.15(open circles) at 3 K under different magnetic fields at (a) 0 T, (b) 6 T and (c) 8.5 T (7 T for $x$=0.15). The horizontal bars indicate the instrumental {\it Q}-resolution. The solid lines are results fitted with a single Gaussian function by convoluting the resolution. 
}
\end{figure}
%
%
\noindent
We note that the magnetic intensity is superposed on the temperature-independent non-magnetic superlattice peak, whose origin was discussed from the viewpoint of lattice distortion\cite{Kurahashi} and/or an impurity phase caused by heat treatments\cite{Matsuura}. 
Measurements with the tighter collimation revealed that the Gaussian line-width is broader than the resolution limited value indicated with the horizontal bar in Fig. 2(a). 
The intrinsic half-width was evaluated to be 0.0049$\pm$0.0008 $\AA$$^{-1}$, corresponding to the inverse of the size of ordered region of 205$\pm$32 $\AA$ in the CuO$_2$ planes. 
These observations suggest that the commensurate short-range AF order coexists or phase-separates with the bulk superconductivity at zero field. 
The peak-width shows no remarkable field dependence, in contrast to the result for LSCO in which the long-ranged AF order is stabilized under the field\cite{Katano,Lake_Nature,Khaykovich1,Khaykovich2}. 
The field-enhanced intensity shows a maximum at $\sim$5 T which is about 25$\%$ of the intensity at 0 T. 
With increasing field beyond 5 T the intensity starts to decrease (Fig. 4) and at $\sim$9 T becomes comparable with that at zero-field. 
No magnetic signal was observed in the $x$=0.15 sample even under magnetic fields. 

In Nd$_{1.85}$Ce$_{0.15}$CuO$_{4}$ studied by Kang and coworkers similar field dependence of the magnetic intensity was observed\cite{Kang}. 
However, the magnetic intensity is much stronger than that of PLCCO, therefore, a strikingly large effect of Nd spins should be included in the field dependence of its magnetic intensity at ($\pi$,$\pi$). 
We further note that a field-induced complicated spin structure appears for Nd$_{1.85}$Ce$_{0.15}$CuO$_{4}$, while no evidence of such a structure was seen in Pr$_{0.89}$LaCe$_{0.11}$CuO$_{4}$. 

The field effect was observed only for the sample with $x$=0.11 where the magnetic intensity exists in zero field. 
This result is comparable with that obtained for LSCO; the static AF order in the SC phase is {\it enhanced} \cite{Katano,Lake_Nature,Khaykovich1,Khaykovich2}, whereas no magnetic order is {\it induced} in the optimally doped or slightly overdoped spin-gap states\cite{Lake_Science}. 
\linebreak
%
\begin{figure}[t]
\centerline{\epsfxsize=2.75in\epsfbox{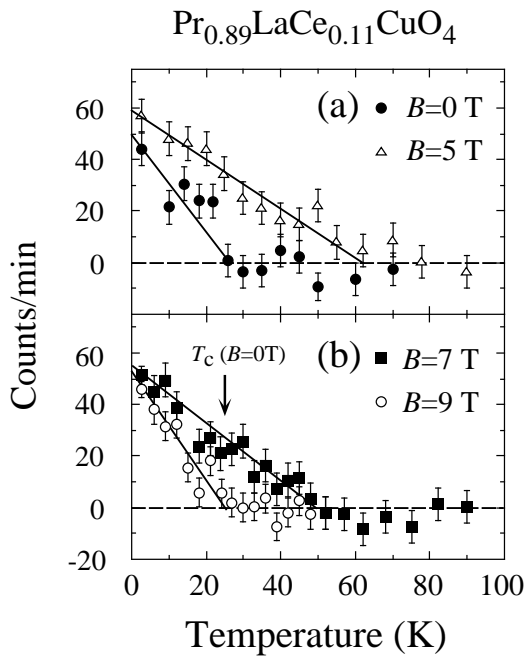}}
\caption
{
Temperature dependence of peak intensity at (1.5 0.5 0) reciprocal-lattice position measured at (a) $B$=0 T, 5 T, (b) 7 T and 9 T for Pr$_{0.89}$LaCe$_{0.11}$CuO$_4$.
}
\end{figure}
%
\noindent
However, in the case of the hole-doped La-214 systems, the intensity increases monotonically with field strength \cite{Lake_Nature,Khaykovich1,Khaykovich2} but no signature of a peak in the field dependence was so far observed.

We next investigated the $T$-dependence of the peak intensity in the $x$=0.11 sample for $B$$\leqslant$9 T. 
As seen in Fig. 3(a), the magnetic intensity appears below $\sim$$T_c$ at zero-field. 
With decreasing temperature, the intensity lineally increases similar to the case of the magnetic order in random systems\cite{Aeppli,Tranquada1}. 
Interestingly, $T_m$ increases up to $\sim$60 K at 5 T and starts to decrease upon further increase of the field. 
Such a field dependence of $T_m$ was not reported for Nd$_{1.85}$Ce$_{0.15}$CuO$_4$ possibly caused by a gradual increase of Nd$^{3+}$ moments in the magnetic order which prevents from a precise determination of $T_m$. 
More interestingly, as shown in Fig. 4, the field dependences of $T_m$ and the peak-intensity at 3 K are quite similar. 
We note that the linear relation between the N\'{e}el temperature and the staggered moment and their large field effect are often seen in the weak itinerant antiferromagnets and theoretically interpreted within the Fermi liquid framework\cite{Moriya}. 
A recent nuclear magnetic resonance study\cite{Zheng} for optimally doped PLCCO also predicted the Fermi-liquid ground state. 
Therefore, the magnetism in the $n$-type cuprate near the AF-SC phase boundary can be regarded as the weak itinerant antiferromagnet, in contrast to that in the hole doped cuprates in which magnetic interaction is robust against doping due to the charge segregation into stripes\cite{Tranquada,Fujita_LBCO}.
\linebreak
%
\begin{figure}[t]
\centerline{\epsfxsize=2.75in\epsfbox{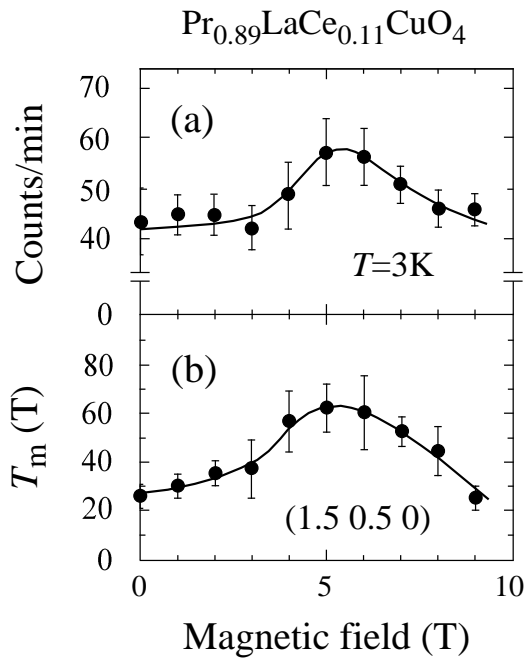}}
\caption
{
Field dependence of (a) the peak-intensity measured at (1.5 0.5 0) at 3 K and (b) the magnetic ordering temperature for Pr$_{0.89}$LaCe$_{0.11}$CuO$_4$.  Solid lines are guide to the eyes.
}
\end{figure}
%
\noindent
The shift of $T_m$ by magnetic field suggests the field dependence of the quantum critical concentration for the magnetic order. 
To confirm this point a neutron scattering study in a narrow concentration range of 0.11$<$$x_c$$\leqslant$0.15 for PLCCO is needed to detect a possible field-induced magnetic order in the samples with no magnetic order at zero field and low temperatures.
%

Now we discuss the relation between the field effect and the superconductivity. 
Systematic $\mu$SR measurements on the PLCCO systems revealed a decrease in volume fraction of the magnetic order, staggered moment and $T_m$ upon electron-doping\cite{Fujita_mSR} near the AF-SC phase boundary. 
This result suggests an existence of AF ordered islands segregated from the SC background in the $x$=0.11 sample. The nonmagnetic 3d$^{10}$ Cu-cites introduced by doped electrons in Cu 3$d$ orbitals may play some role for the formation of the AF islands\cite{Yamada_NCCO,Zheng}. 
In this situation, magnetic field affects both SC and AF ordered regions. 
In the present case, since the maximum value of $T_m$ under magnetic field exceeds $T_c$($B$=0 T), we speculate that the field effect on the AF islands is important to understand the effect on the spin correlation. 
Absence of field-induced AF order in the SC region may relate with an energy gap in magnetic excitation as seen in NCCO\cite{Yamada_NCCO}. 
On the other hand, the field-induced AF order around SC vortices is predicted for both hole- and electron-doped cuprates\cite{Chen}. 
We note that the upper critical field of optimally-doped PLCCO is comparable to the value at which the field dependence of both magnetic intensity at $T$=3 K and $T_m$ show a peak\cite{Zheng}. 
Thus, the origin of the simultaneous suppression of magnetic intensity and $T_m$ at high fields beyond 5 T is important to understand the magnetic field effect on high-$T_c$ cuprates. 

In conclusion, we have investigated the effect of magnetic field on the static spin correlations in the electron-doped superconductors Pr$_{1-x}$LaCe$_{x}$CuO$_4$ with $x$=0.11 and 0.15 by neutron scattering measurements. 
The field dependences of the magnetic intensity at low temperatures and the onset temperature of commensurate AF order are similar and show a peak at $\sim$5 T in the $x$=0.11 sample. 
In contrast, a static AF order was not induced in the $x$=0.15 sample under the fields up to 8.5 T. 
These new findings of field-effect in the electron-doped system nearly free from the effect of rare-earth spins would provide important clues for understanding the common mechanism of high-$T_c$ superconductivity in the doped antiferromagnets. 

We gratefully thank R. Kadono for invaluable discussions and Y. Shimojyo for technical assistance. This work was supported in part by the Japanese Ministry of Education, Culture, Sports, Science and Technology, Grant-in-Aid for Scientific Research on Priority Areas (Novel Quantum Phenomena in Transition Metal Oxides), for Scientific Research (B), for Encouragement of Young Scientists, and for Creative Scientific Research "Collaboratory on Electron Correlations - Toward a New Research Network between Physics and Chemistry". 
%


\end{document}